\documentclass[12pt]{article}
\usepackage[dvips]{epsfig}
\usepackage{a4wide}

\newcommand{\beq}{\begin{equation}}
\newcommand{\eeq}{\end{equation}}
\newcommand{\bea}{\begin{eqnarray}}
\newcommand{\eea}{\end{eqnarray}}
\newcommand{\lab}[1]{\label{#1}}

\newcommand{\e}{\mbox{e}}

\newcommand{\rh}{\rho}

\newcommand{\ph}{\varphi}
\newcommand{\om}{\omega}
\newcommand{\la}{\lambda}
\newcommand{\G}{\Gamma}

\newcommand{\ks}{\xi}

\newcommand{\bk}{{\bf k}}
\newcommand{\bx}{{\bf x}}
\newcommand{\by}{{\bf y}}

\newcommand{\bu}{{\bf u}}
\newcommand{\bv}{{\bf v}}
\newcommand{\bw}{{\bf w}}

\newcommand{\cO}{{\cal O}}

\newcommand{\nn}{\nonumber}

\newcommand{\9}{\partial}
\newcommand{\de}{\delta}

\newcommand{\lb}{\lbrace}
\newcommand{\rb}{\rbrace}
\newcommand{\cdo}{\hspace{-0.05 em} \cdot \hspace{-0.05 em} }  

\newcommand{\sect}[1]{\bigskip \bigskip \noindent{\Large #1 \\}\par }
\newcommand{\tr}{\mbox{tr}}

\begin{document}
\rightline{TUW-97-06}
\rightline{hep-th/9703018}

\bigskip \bigskip \bigskip \bigskip

\begin{center} 
{\Huge
The Quantum Liouville Equation \\ \medskip for the Effective Action }

\bigskip \bigskip \bigskip \bigskip

Herbert Nachbagauer\footnote{email: herby \@@ tph16.tuwien.ac.at}\\

\bigskip 

Institut f\"ur Theoretische Physik \\
Technische Universit\"at Wien \\ 
Wiedner Hauptstra{\ss}e 8-10 \\
A-1040 Wien, Austria \\
\end{center} 

\bigskip \bigskip 

\abstract{Starting from the von~Neumann equation, we construct the 
quantum evolution equation for the effective action for systems 
in mixed states. This allows us to find the hierarchy of 
equations which describe the time evolution of equal time correlators.
The method is applied explicitly to a scalar theory with quartic 
self-interaction.\\ 
\smallskip

 } 

\newpage

\sect{Introduction}
During the last years, great progress has been made in the study of 
statistical systems of quantized fields in thermal equilibrium.
Many physically interesting quantities, however, are typically 
related to out-of equilibrium scenarios, like density 
fluctuations in the 
early universe, the radiation signature of hot quark-gluon
matter or baryon number violating processes. The study of such 
quantities in systems which are not near to equilibrium 
requires a framework which is able to deal with 
mixed states characterized by general density matrices. 

Although the Schr\"odinger equation already determines the exact  
evolution of initial density matrices and can be integrated to solve 
the problem of time evolution, it is of less use for practical
purposes. What one is interested in and what is used to describe 
a physical system is not the density matrix 
but a number of observables derived from it. 
At first sight it may appear to be rather a matter of taste to 
study time evolution of observables instead of the density matrix,
but there is also a profound physical reason to change the point of 
view. Time integration of the density matrix requires the exact 
knowledge of the initial state, which is not accessible since 
observation is usually restricted to only a small number data. 
Consequently, the description of a system requires apart from the 
knowledge of the exact motion of observables also a statistical 
assumption replacing the incomplete knowledge of the initial state. 

In this letter we focus on the first step of the problem.
We investigate on the exact counterpart of what the Schr\"odinger 
equation is for the density matrix: A quantum Liouville equation for the 
effective action which encodes in a compact form the time evolution  
of all equal time correlators being one basis of a complete set of 
observables. 

\sect{Construction} 
The following construction will be formulated in the Schr\"odinger picture, 
that means operators are time independent and time evolution 
is contained in the density matrix. The resulting evolution equation,
however, is independent of the picture chosen.

Starting point is the von~Neumann equation for the density
matrix operator,
\beq 
i \9_t \rho (t)  = [ H , \rho (t) ] , \lab{vn} 
\eeq 
with Hamiltonian $H$. It is a functional of the position operator 
$\Phi ( \bx ) $ and its canonical momentum operator $\Pi (\bx) $. 
In a given mixed state described by a density matrix $\rho(t) $, operator 
expectation values are given by the trace average
\beq 
\left< \cO \right>_t = \tr \left( \cO \rh ( t ) \right) ,
\eeq 
where the hermitian density matrix is normalized to unity, $\tr \rh (t) =1 $,
which is consistent with unitary time evolution.
 
We introduce the generating functional of polynomials in 
momentum and position operators by 
\beq 
Z[j,k;t] = \left< D[j,k] \right>_t  = \tr ( D[j,k] \rh(t) ) 
\eeq 
with current operator  
\beq 
D[j,k] = \exp ( i j \cdo \Phi + i k \cdo \Pi ) 
\eeq 
and shorthand notation $j \cdo \Phi := \int d^3\!x j(\bx) \Phi(\bx) $.

Since neither position nor momentum operators explicitly depend on
time, integration is done with respect to space only, and also
there is no need to introduce time dependent currents. This is a very 
special feature of the Schr\"odiger picture where all the dynamics
is hidden in the states and the density matrix.
The definition breaks Lorentz invariance. However, it is broken anyway
since in the Hamilton formulation the time direction (evolution
direction) orthogonal to the time slice of quantisation plays a 
privileged r\^ole.  

The current operator can be rewritten as 
\beq 
D[j,k] =  \exp ( i j \cdo \Phi ) \exp ( i k \cdo \Pi ) 
           \exp ( \frac{i}{2} j \cdo k )   = 
         \exp ( i k \cdo \Pi ) \exp ( i j \cdo \Phi ) 
           \exp ( -\frac{i}{2} j \cdo k )  
\eeq
using the fundamental commutator $ [ \Phi(\bx) ,\Pi(\by) ] = 
i \delta^3 (\bx - \by ) $  and the first non-trivial term in the
Campbell-Baker-Hausdorff formula. 
From that representation, powers of position and momentum operators can 
be obtained by the functional differential operators, 
\beq
\left( \frac{\de}{i \de j} - \frac{1}{2} k \right) D[j,k] = \Phi D[j,k], \qquad 
\left( \frac{\de}{i \de k} + \frac{1}{2} j \right) D[j,k] = \Pi D[j,k],
\lab{direl} 
\eeq
or, equivalently, by acting from the left with
\beq
D[j,k] \left( \frac{\stackrel{\leftarrow}{\delta} }{i \de j} + 
\frac{1}{2} k \right) =  D[j,k] \Phi ,\qquad 
D[j,k] \left( \frac{\stackrel{\leftarrow}{\de}} {i \de k} - \frac{1}{2} j 
\right) =  D[j,k] \Pi .
\lab{direl2} 
\eeq
Polynomial operator expectation values $\cO[ \Phi,\Pi ]$ can thus be 
cast by firstly constructing the operator with inverse order 
by reading it from the right to the left, secondly the replacements 
$  \Phi  \to ( \de / i \de j - k/2 ) $ and 
$ \Pi \to ( \de / i \de k +  j / 2 ) $ in the inverse ordered
operator, third acting with this expression on the generating functional, 
and finally letting $j,k \to 0 $ afterwards. 
We note that the functional differential operator replacements have 
the negative commutator with respect to their field operator counterparts. 
We also note that if we used the representations  
$  \Phi  \to  \de / i \de j  $ and 
$ \Pi \to \de / i \de k  $ instead of those of Eq.\ (\ref{direl}), 
we get the symmetrically ordered operator polynomial.

Let us construct the time evolution of the generating 
functional. From its definition and the equation of motion for the density
matrix, we get 
\beq 
\9_t Z[j,k;t] = i\, \tr \left( H[\Pi,\Phi] D[j,k] \rh(t) \right) - 
i \, \tr \left( D[j,k] H[\Pi,\Phi] \rh(t) \right) .
\eeq
In view of the arguments given above, the Hamiltonian can now be replaced 
by a functional differential operator which can be drawn out of the 
trace. We get a Liouville equation for the generating functional
\beq
\9_t Z[j,k;t] = L[j,k] Z[j,k;t]  \lab{eomz}
\eeq
with Liouvillian
\beq 
L[j,k] = i H^{\mbox{inverse}} [\frac{\de}{i \de k} + \frac{1}{2} j ,
 \frac{\de}{i \de j} - \frac{1}{2} k ] -
i H[\frac{\de}{i \de k} - \frac{1}{2} j,
  \frac{\de}{i \de j} + \frac{1}{2} k ]. \lab{liou} 
\eeq
It is interesting to restore powers of $\hbar$ at this stage. We note 
that the terms with factors one-half come from the fundamental 
commutator and are of relative order $\hbar$. Expanding the r.h.s.\ 
in $\hbar$, the leading term cancels due to the relative minus sign in
the summands in (\ref{liou}). The first non-vanishing contribution is 
linear in $\hbar$ as the time derivative which 
also contains one $\hbar$. That contributions correspond to the 
classical limit of the Liouville equation. 

$ Z[j,k;t]$ is the generating functional of all equal-time correlators
corresponding to  Green functions at equal time.  
The generating functional of the connected correlators $W[j,k;t]$, 
however, is related to $Z[j,k;t]$ by 
\beq 
\e^{W[j,k;t]} =  Z[j,k;t] .
\eeq
This relation implies for the r.h.s.\ of the Liouville equation
\beq
L[j,k] \e^{W[j,k;t]} = \bar L[ W ] [j,k;t]   \e^{W[j,k;t]} 
\eeq
which defines a non-linear evolution operator. The exponential also
appears in the time derivative of the Liouville equation 
and thus cancels out. Evolution is now given by the Liouvillian--like 
equation
\beq 
\9_t W[j,k;t] = \bar L[W] [j,k;t] \lab{eomw} 
\eeq
This equation qualitatively differs from the 
counterpart for $Z[j,k;t] $ since the operator $\bar L$ is
non-linear in its action on $W$. 

In order to establish a relation to observables, 
we introduce the classical c-number field and momentum 
variables. We define the quantities
\beq 
\ph(t,\bx)  := \frac{\de W[j,k;t]}{i\de j(\bx)} ,\qquad 
\pi (t,\bx)  := \frac{\de W[j,k;t]}{i\de k(\bx)} \lab{class}
\eeq
which coincide with the expectation values $\left<\Phi(\bx) \right>_t$ and 
$ \left<\Pi (\bx)\right>_t $ when the currents vanish. We want to point out 
that the classical variables $\ph(t,\bx)|_{j=k=0} $ and  
$\pi (t,\bx)|_{j=k=0} $ do depend on time and are independent of the 
picture chosen -- in the opposite to their Schr\"odinger  operator 
counterparts. In the definition  (\ref{class}), the classical
field and momentum are functionals of the currents $j,k$. 
Suppose we can invert the relation and express the 
currents as functionals $ j = j[\ph,\pi],\, k=k[\ph,\pi]$, then 
the effective action defined by 
\beq 
\G [\ph,\pi;t] =  \ph(t) \cdo j + \pi(t) \cdo k + i W[j,k;t] 
\lab{gdef} 
\eeq 
can be expressed as a functional of the classical variables. 
It obeys a relation dual to the definition (\ref{class})
\beq  
 \frac{\de \G[\ph(t),\pi(t);t]}{\de \ph (t)} = j, \qquad 
 \frac{\de \G[\ph(t),\pi(t);t]}{\de \pi (t)} = k . \lab{dual} 
\eeq 
That implies that the currents vanish at the minimum of the effective action 
where the classical quantities have their 
interpretation as operator expectation values. 

In order to apply terms correctly, let us remind the reader 
that the quantity $\G$ discussed here differs slightly from the 
definition used in standard field theory, namely, it 
corresponds to the generating functional of vertex functions at 
equal times. 
Here, the effective action itself has explicit time dependence from
the classical fields, and an implicit one  
inferred from the evolution of the density matrix.
The total time derivative thus behaves as 
$d\G / dt =\dot \ph \cdo \de \G / \de \ph  + \dot \pi \cdo \de \G / \de \pi
  +  \9_t \G $, 
which, when compared with the time derivative of the 
definition (\ref{gdef}), implies
\beq 
\9_t \G[\ph,\pi ;t] = i \9_t W [j,k; t ]
\eeq
where we took into account the time independence of the currents 
and the relation (\ref{dual}).  
It is now straight-forward to translate the equation of motion for 
the generating functional of the connected Green functions into
an equation for $\G$. 
One only has to express the Liouvillian in the   
evolution equation (\ref{eomw}) in terms of $\ph$ and $\pi$. 
In $\bar L[W]$, currents can be replaced by the  
relations (\ref{dual}).  The first derivatives of $W$ are given in 
Eqs.\ (\ref{class}), and higher derivatives can be found by successive 
differentiation of that relation. We get for the first few terms 
\bea
-i W_{,jj} &=& \ph_{,j} = \G_{,\ph\ph}^{-1} =: G^{(\ph\ph)}  \nn \\
-i W_{,jjj} &=& - \G_{,\ph\ph}^{-2} \left( 
\G_{,\ph\ph\ph} \G_{,\ph\ph}^{-1} + 
\G_{,\ph\ph\pi} \G_{,\ph\pi}^{-1} \right) \nn \\
-i  W_{,jk} &=& \ph_{,k} = \pi_{,j} = \G_{,\ph\pi}^{-1} =: G^{(\ph\pi)} \nn \\
-i W_{,jjk} &=& - \G_{,\ph\ph}^{-2} \left( 
\G_{,\ph\ph\ph} \G_{,\ph\pi}^{-1} + 
\G_{,\ph\ph\pi} \G_{,\pi\pi}^{-1} \right) \nn \\
-i W_{,jkk} &=& - \G_{,\ph\pi}^{-2} \left( 
\G_{,\ph\ph\pi} \G_{,\ph\pi}^{-1} + 
\G_{,\ph\pi\pi} \G_{,\pi\pi}^{-1} \right) 
\lab{mom} 
\eea
where we wrote shorthand $\de \G/\de \ph =: \G_{,\ph}$ etc., and 
$W$ itself has to be replaced by 
$-i ( \G - \ph \cdo \G_{,\ph} - \pi \cdo \G_{,\pi} )$.
Along that line of arguments, we can translate the operator
$\bar L [W] $ and  
finally get the desired Liouville equation for the effective action,
to wit
\beq
\9_t \G =  F[\G] = i \left. \bar L [W ] \right|_{j[\ph,\pi],k[\ph,\pi]} .
\lab{eomga}
\eeq
The classical limit of it was studied in \cite{wett} using a 
different approach. 
The relation (\ref{eomga}) is local in the time parameter and the determining
equation for $\G$. Expanding the effective action in powers of 
$\ph$ and $\pi$
\bea  
\G[\ph,\pi] &= & 
\sum_{n=1}^{\infty} \frac{1}{n! }  
\left( \prod_{i=1}^n \int d^3 \bx_i \right) \times  \nn \\ 
&& \sum_{j=0}^n 
\G_n^j (\bx_1\ldots \bx_n;t ) \ph(\bx_1,t) \ldots \ph(\bx_j,t) 
\pi(\bx_{j+1} ,t) \ldots \pi(\bx_n,t )  \lab{gn}
\eea
and comparing the corresponding coefficients at both sides of 
(\ref{eomga}), one gets 
an infinite non-linear and highly coupled hierarchy for the time 
dependence of the proper vertex functions $\G_n^j$.
The solution involves an infinite number of initial conditions 
corresponding to the infinite set of observables necessary to 
describe a system completely. Suppose we can find a solution of
that hierarchy, at least a perturbative one, one may ask for the 
time evolution of observables. 
We know that the c-number fields $\ph$ and $\pi$ have their 
interpretation as expectation values of the position and momentum 
operator when the currents vanish. Thus, from the dual relation 
(\ref{dual}) we find the effective equations of motion
\beq 
 \frac{\de \G[\ph(t),\pi(t);t]}{\de \ph (t)} = 0, \qquad 
 \frac{\de \G[\ph(t),\pi(t);t]}{\de \pi (t)} = 0 . \lab{eomeff} 
\eeq 
as variational minimum of the effective action. These equations determine 
the time evolution of the average field $\left< \Phi \right>_t $ and momentum 
$\left<\Pi \right>_t $ resp. Expectation values of symmetrically ordered 
higher order operators are given by the 
derivatives of $W$ in Eq.\ (\ref{mom}) evaluated at the solution of
(\ref{eomeff}).

\sect{Scalar field}
Let us consider the non-interacting motion of a scalar field 
characterized by the Hamiltonian
\beq 
H^0 = \frac{1}{2} \int d^3\! x \left( \Pi^2 + 
\Phi ( - \Delta + m^2 ) \Phi \right)
\eeq 
which, by Eq.\ (\ref{liou}), gives rise to the free Liouvillian
\beq 
L^0 = i \int d^3\!x \left( j(\bx)  \frac{\de}{i \de k(\bx) } -
 k(\bx)(- \Delta + m^2 ) \frac{\de}{i \de j(\bx) } \right).
\eeq 
Since, in this simple case, the operator  is even linear in the 
functional derivatives, its action on the exponential is also linear
in $W$, $L^0 \e^W = ( L^0 W ) \e^W $. Rewriting everything 
in terms of the effective action yields
\beq 
\9_t \G =  F^0[\G] =  \int d^3 \!x \left( \ph(\bx) ( - \Delta + m^2 ) 
\frac{\de \G}{\de \pi(\bx) } -
\pi (\bx)  \frac{\de \G }{ \de \ph(\bx) } \right) \lab{free}
\eeq
which has a remarkably 
homogeneity property. The r.h.s.\  is a map of the class of 
polynomials in $\ph$ and $\pi$ of a given fixed order onto itself and
the n-point vertices $\G_n $ in Eq.\ (\ref{gn}) obey a closed equation 
each. Time evolution can best be made explicit by a diagonalisation 
procedure using the normal coordinates $\ks,\ks^*$
\beq 
\ks = \sqrt{\frac{\om}{2} }\ph + \frac{i}{\sqrt{2 \om} }\pi,  \qquad 
\om_\bk =  \sqrt{m^2 + \bk^2},
\eeq
where the expansion coefficients in (\ref{gn}) transform accordingly. 
The Liouville operator now takes the form 
\beq 
F^0[\G] = i \int d^3\! k  \, \om \left( 
\ks \frac{\de \G[\ks,\ks^*]}{\de \ks} -  
\ks^* \frac{\de \G[\ks,\ks^*]}{\de \ks^*}  \right) 
\eeq
which implies for time evolution (the index $j$ now refers to the number 
of $\ks$'s in the expansion in normal coordinates)
\beq
{\G_n^j}^{\lb \ks \rb} 
(\bk_1 \ldots \bk_n;t)  = {\G_n^j}^{\lb \ks \rb}  (\bk_1 \ldots \bk_n;0) 
\exp ( i \sum_{i=1}^j \om_{\bk_i}  t  - i  \sum_{i=j+1}^n \om_{\bk_i} t ).
\eeq
As expected, each proper mode just oscillates harmonically. We want to
point out that motion is perfectly compatible with arbitrarily high 
order vertex functions, and there is no physical argument to neglect 
them in the first place. In that sense, in a non-interacting system 
higher correlation functions have to be taken into account in order to 
allow for general initial conditions.

In contrast to the simple form of the time evolution of the 
effective action, 
the solution of the on-shell conditions (\ref{eomeff}) 
involves to resolve a non-local non-linear equation for the 
classical fields $\ph$ and $\pi$ even in the non-interacting theory.        

The inclusion of a quartic interaction term 
$H = H^0 + \la \int d^3\!x \ph^4 $ modifies the Liouvillian by 
\beq 
L^{(\la)} = -i \la \int d^3\! x \left( 4 
\left( \frac{\de}{i\de j(\bx) } \right)^3 k(\bx) + 
\frac{\de}{i\de j(\bx) }  k^3(\bx) \right).
\eeq    
This gives rise to the non-linear Liouvillian 
\bea    
F^\la [\G] &= &\la \int d^3 \! x \left( 
4 \frac{\de\G}{\de \pi(\bx) } \left[ 
\int d^3 \!u\int d^3\! v \int d^3 \!w G^{(\ph\ph)} (\bx,\bu) 
G^{(\ph\ph)}(\bx,\bv) \times \right. \right. \nn \\ 
&& 
\left. \left. \left[ G^{(\ph\ph)}(\bx,\bw) 
\frac{\de^3\G}{\de \ph(\bu)\de \ph(\bv)\de \ph(\bw)}
+ G^{(\ph\pi)} ( \bx,\bw) 
\frac{\de^3\G}{\de \ph(\bu)\de \ph(\bv)\de \pi(\bw)}
\right] \right. \right. \nn \\ 
&& 
\left. \left. 
-3 i \ph(\bx) G^{(\ph\ph)}(\bx,\bx) + \ph^3(\bx) \right] +
\left( \frac{\de\G}{\de \pi(\bx) } \right)^3 \ph(\bx) 
\right). 
\eea
The graphical representation of the complete time evolution 
including the free part is depicted in Fig.\ 1.
The filled blob denotes the generating functional $\G$, and full 
and dashed legs attached to it correspond to 
derivatives with respect to $\ph$ and $\pi$ respectively. The 
unfilled circle with 
one leg stands for a factor $\ph$ and with two legs for the two-point
correlators $G^{(\ph\ph)}$ and $G^{(\ph\pi)}$. The diamond stands for 
the free kernel operator $-\Delta + m^2 $ and the dashed circle for a 
factor $\pi$.

The first two graphs correspond to the classical non-interacting 
and the third through sixth graph to the classical interacting 
Liouvillian.  Only the last graph is a quantum correction 
being of relative order $\hbar^2$. The system can be expanded in powers of 
$\ph$ and $\pi$ which leads to a hierarchy of coupled equations for the 
vertex functions $\G_n^j$. 

One may now start to attempt to solve the 
hierarchy for the vertex functions by cutting it 
at a certain order. However, this somehow pragmatic
approach, however, is not guaranteed to yield a physically sensible 
result. Firstly, one runs into consistency problems with the hierarchy. 
Even worse, a check would require to calculate
those terms which have been dropped in the cutting procedure in order
to justify them to be neglected. Moreover, guided by the exact solution 
for the free case, 
where we have already seen that higher correlators 
have to be treated at the same footing as all others, it is at first 
sight not evident why just the more complex dynamical structure 
should admit any such simplification. 
Secondly, cutting the hierarchy implicitly implies an assumption about the 
initial conditions for the higher vertex functions. It is not easy to 
see what physical initial preparation cutting the hierarchy and 
assuming particular initial values just for the lowest vertex 
functions represent. It is even more difficult to figure 
out what statistical assumptions about the system are hidden in 
a particular cutting procedure. 

To summarize, the Liouville equation derived here makes a statement about 
the exact evolution of a system in mixed states. Any statement about the 
thermodynamic properties of that system, and in particular about 
how it approaches an equilibrium state cannot be read of from the 
dynamics but requires the application of the sophisticated tools 
of statistical mechanics.  This goes beyond the scope of this letter
and will be presented elsewhere.  \\

\bigskip \bigskip 

\noindent {\bf Acknowledgement.} 
I thank Christof Wetterich for helpful discussion.
This work is supported by the {\em \"Osterreichische Nationalbank}
under project No.\ 5986.

\newpage 

\begin{figure}        
\begin{center}
\epsfxsize5in
\epsfbox{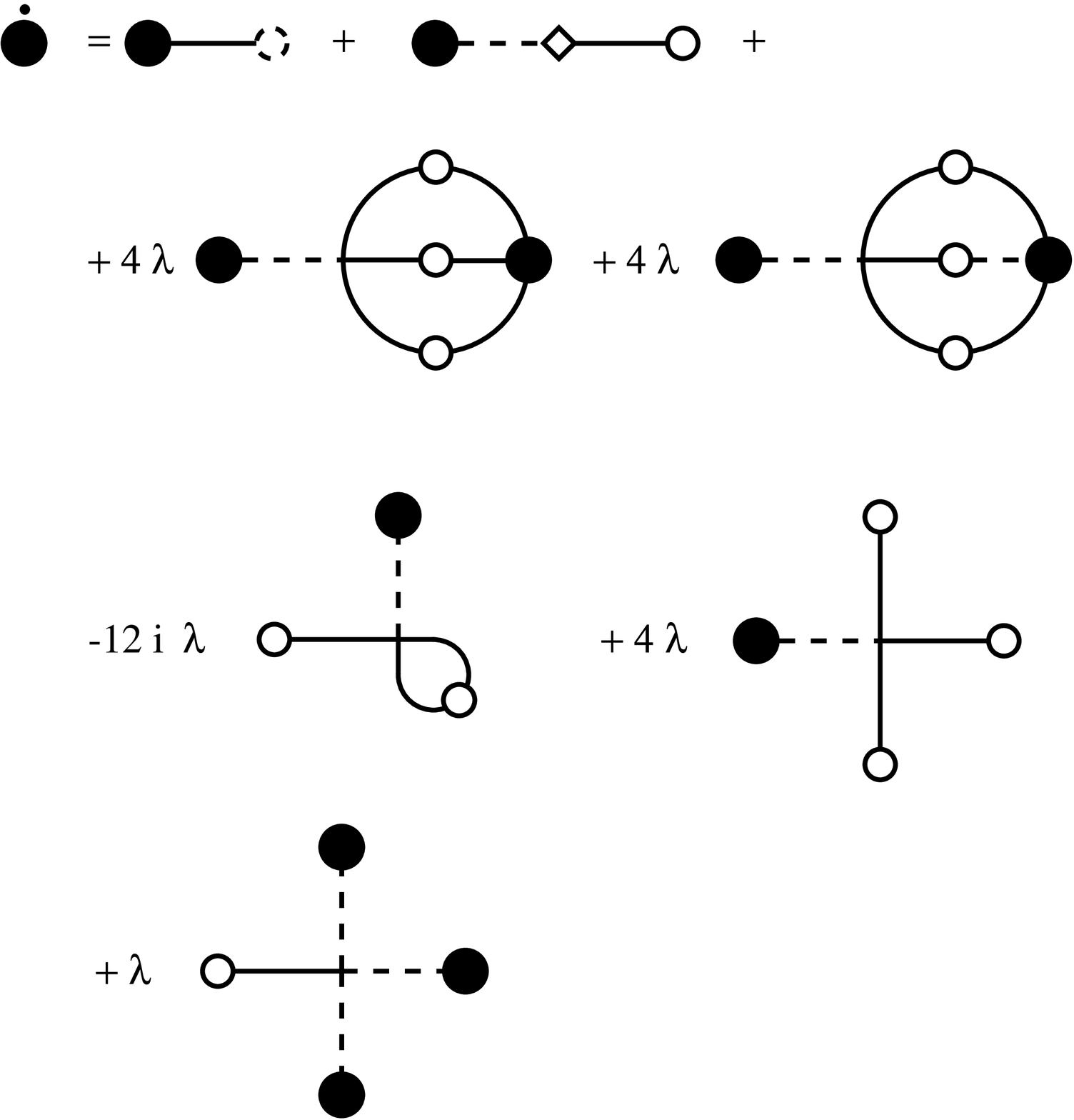}
\caption{Time evolution of the effective action for scalar $\Phi^4$-theory.}
\end{center}                            
\end{figure} 

\end{document}